\documentclass[10pt, a4paper]{article}
\usepackage{geometry}                
\usepackage{graphicx}
\usepackage{amssymb}
\usepackage{epstopdf}
\usepackage{amsthm}
\usepackage{graphicx,psfrag}
\usepackage{amsmath}
\usepackage{enumerate}
\usepackage[small]{caption}
\usepackage{floatrow}
\usepackage{multirow}
\usepackage[usenames,dvipsnames,table]{xcolor}
\begin{document}

\title{A Comparison of \\ the LVDP and $\Lambda$CDM Cosmological Models}
\author{\"{O}zg\"{u}r Akarsu\footnote{E-Mail: oakarsu@ku.edu.tr}  ,  Tekin Dereli\footnote{E-Mail: tdereli@ku.edu.tr}}

\date{}

\maketitle
\begin{center}
\vskip-1cm
\textit{Department of Physics, Ko\c{c} University, 34450 {\.I}stanbul/Turkey.}
\end{center}

\begin{abstract}
We compare the cosmological kinematics obtained via our law of
linearly varying deceleration parameter (LVDP) with the kinematics
obtained in the $\Lambda$CDM model. We  show that the LVDP model is
almost indistinguishable from the $\Lambda$CDM model up to the near
future of our universe as far as the current observations are
concerned, though their predictions differ tremendously into the far
future.
\begin{flushleft}
\end{flushleft}
\end{abstract}

\bigskip

In a  recent paper (Ref. \cite{LVDP12}), we proposed a special law
(LVDP) for the deceleration parameter  $q=-kt+m-1$ that is linear in
cosmic time $t$, where $k>0$ is a constant with the dimensions of inverse time and $m>1$ is a dimensionless constant. This law allows
us to generalize many exact cosmological solutions that one finds in
the literature with a constant deceleration parameter ($q=m-1$, see
\cite{Berman83}), so as to obtain an expansion history of the
universe that fits better with the observations. For instance, with
the choice $k=0.097$ and $m=1.6$ in Ref. \cite{LVDP12}, we set
$q=-0.73$ \cite{Cunha09} for the present universe (13.7 Gyr old) and predict that
the transition from the decelerating to accelerating expansion
should occur at $t_{\rm t}\cong 6.2$ Gyr and at cosmic redshift
value $z_{\rm t}\cong 0.5$. Both of these values are consistent with
current cosmological data.

The standard $\Lambda$CDM cosmological model is the simplest and
arguably the one that most successfully describes the evolution of
the observed universe. However, it suffers from important conceptual
problems related with the presence of a cosmological constant.
Besides that, the analyses of the cosmological
data not only suggest an equation of state (EoS) parameter value
$w\sim -1$ for the dark energy component of the universe but also do
not exclude a time dependent EoS parameter that can pass below the
phantom divide line ($w=-1$), i.e., the quintom dark energy and
hence a Big Rip, in the future of the universe \cite{Cai10}. Nevertheless,
$\Lambda$CDM is still considered a reference cosmological model such
that any viable model should exhibit  similar kinematics with the
$\Lambda$CDM model up to the present age of the universe,
independent of what it would predict for the future kinematics of
the universe. Therefore, in this letter, we compare LVDP and
$\Lambda$CDM models, and show that they cannot be distinguished
observationally today, but would differ tremendously in the
relatively far future. LVDP model predicts that the universe
eventually enters into the super-exponential expansion phase and
ends with a Big Rip while $\Lambda$CDM model predicts that the
universe monotonically approaches the de Sitter universe. We compare
the behavior of the scale factors, Hubble and deceleration
parameters of the $\Lambda$CDM and LVDP models. We
also compare the jerk and the snap parameters that involve,
respectively, the third and the fourth derivatives of the scale
factors. The jerk and snap parameters were not given before when the
LVDP ansatz was first proposed in Ref. \cite{LVDP12}.

\bigskip

We describe our observed universe by the Friedmann-Lemaitre-Robertson-Walker (FLRW) metric
\begin{eqnarray}
\label{eqn:metric}
ds^2=-dt^2+a^2(t)\,\frac{dx^{2}+dy^{2}+dz^{2}}{\left[1+\frac{\kappa}{4}(x^{2}+y^{2}+z^{2})\right]^2},
\end{eqnarray}
where $a(t)$ is the cosmic scale factor and the spatial curvature
index $\kappa=-1,\,0,\,1$ corresponds to spatially open, flat and
closed universes, respectively. We introduce four cosmological
parameters that describe the kinematics of the universe, namely the
 Hubble parameter and three (dimensionless) parameters; the deceleration,
jerk and snap parameters given, respectively, as follows:
\begin{eqnarray}
H=\frac{\dot{a}}{a},\quad q=-\frac{\ddot{a}}{aH^2},\quad j=\frac{\dddot{a}}{aH^3},\quad s=\frac{\ddddot{a}}{a H^4},
\end{eqnarray}
where an overdot denotes ${\rm d}/{\rm d}t$ (see \cite{Sahni03,Visser04,Dunajski08}).

\medskip

Einstein's field equations are solved in the standard $\Lambda$CDM
model for a mixture of pressure-less matter (including cold dark
matter, CDM) and a cosmological constant $\Lambda$ with positive
sign for the spatially flat space-time ($\kappa=0$). The kinematics
for  $\Lambda$CDM model follows \cite{Sahni00,GronHervik}:
\begin{eqnarray}
&a_{\rm \Lambda CDM}=a_{1}\,\sinh^{\frac{2}{3}}\left(\sqrt{\frac{3\Lambda}{4}}\,t\right),\quad H_{\rm \Lambda CDM}=
\sqrt{\frac{\Lambda}{3}}\coth\left(\sqrt{\frac{3\Lambda}{4}}\,t\right),&\\
\nonumber
&\quad q_{\rm \Lambda CDM}=\frac{1}{2}-\frac{3}{2}\tanh^{2}\left(\sqrt{\frac{3\Lambda}{4}}\,t\right),\quad j_{\rm \Lambda CDM}=
1\quad \textnormal{and}\quad s_{\rm \Lambda CDM}=-\frac{7}{2}+\frac{9}{2}\tanh^{2}\left(\sqrt{\frac{3\Lambda}{4}}\,t\right).&
\end{eqnarray}

\medskip

In the LVDP model on the other hand, we introduce the LVDP ansatz
$q=-kt+m-1$ from the beginning and obtain the effective
energy-momentum tensor by substituting the corresponding scale
factor into the Einstein's field equations, rather than introducing
first the matter fields (see \cite{LVDP12} for details). We find
\begin{eqnarray}
\label{eqn:sf}
&a_{\rm LVDP}=a_{2}\,e^{\frac{2}{m}{\rm{arctanh}}\left(\frac{k}{m}t-1\right)},\quad H_{\rm LVDP}=-\frac{2}{t(kt-2m)},\quad q_{\rm LVDP}=-kt+m-1,&\\
\nonumber
&j_{\rm LVDP}=\frac{3}{2}\,{k}^{2}{t}^{2}-3 k \left( m-1 \right)\,t+(2m-1)(m-1)&
\\
\nonumber
\textnormal{and}
\\
\nonumber
&s_{\rm LVDP}=3\,{k}^{3}{t}^{3}-9{k}^{2} \left( m-1 \right)\,{t}^{2}+6k\, \left( 2m-1 \right)\left( m-1 \right)\, t-6\,{m}^{3}+11\,{m}^{2}-6\,m+1.&
\end{eqnarray}
We would like to note here that the kinematics of the LVDP model can
be generated from a constant such that $k=-\dot{q}$, while the
kinematics of the $\Lambda$CDM could be generated from the jerk
parameter value $j=1$ \cite{Sahni03,Visser04,Dunajski08}.

\medskip

\begin{figure}[ht]
\floatbox[{\capbeside\thisfloatsetup{capbesideposition={left,top},capbesidewidth=0.35\textwidth}}]{figure}[\FBwidth]
{\caption{Scale factors $a$ versus cosmic time $t$ for the LVDP (solid) and $\Lambda$CDM (dashed) models. The vertical line represents the present time of the universe $13.7$ (Gyr). The scale factor diverges at $t=34.992$ (Gyr), i.e., Big Rip occurs, in the LVDP model.}\label{fig:sf}}
{\includegraphics[width=0.6\textwidth]{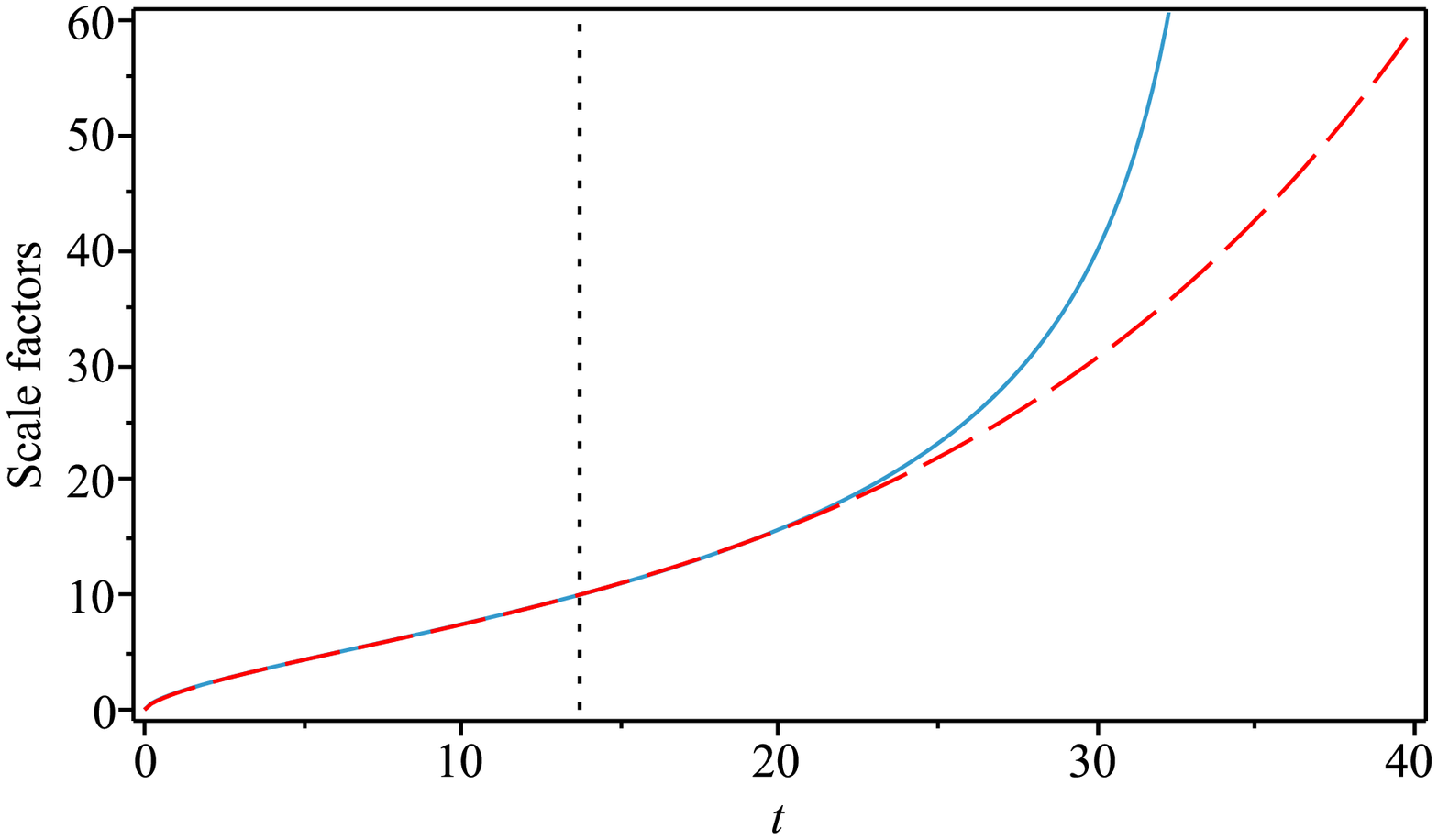}}
\end{figure}

\begin{figure}[ht]
\floatbox[{\capbeside\thisfloatsetup{capbesideposition={left,top},capbesidewidth=0.35\textwidth}}]{figure}[\FBwidth]
{\caption{Hubble parameters $H$ versus cosmic time $t$ for the LVDP (solid) and $\Lambda$CDM (dashed) models. The vertical line represents the present time of the universe $13.7$ (Gyr). The Hubble parameter diverges at the Big Rip time $t=34.992$ (Gyr) in the LVDP model.}\label{fig:h}}
{\includegraphics[width=0.6\textwidth]{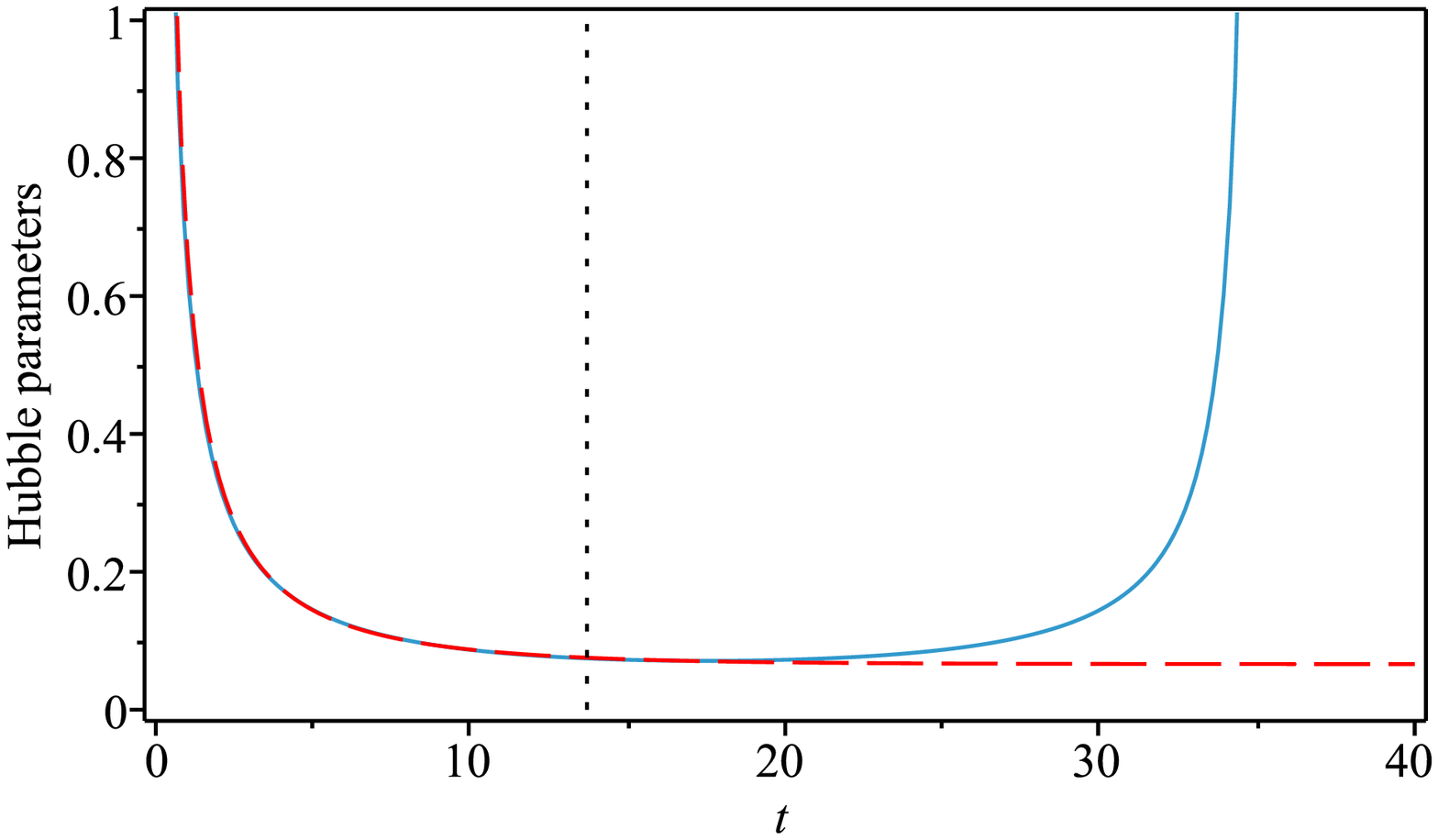}}
\end{figure}

\begin{figure}[ht]
\floatbox[{\capbeside\thisfloatsetup{capbesideposition={left,top},capbesidewidth=0.35\textwidth}}]{figure}[\FBwidth]
{\caption{Deceleration parameters $q$ versus cosmic time $t$ for the LVDP (solid) and $\Lambda$CDM (dashed) models. The vertical line represents the present time of the universe $13.7$ (Gyr). The line of the deceleration parameter in the LVDP model ends at $t=34.992$ (Gyr) since the universe ends with a Big Rip at this time.}\label{fig:q}}
{\includegraphics[width=0.6\textwidth]{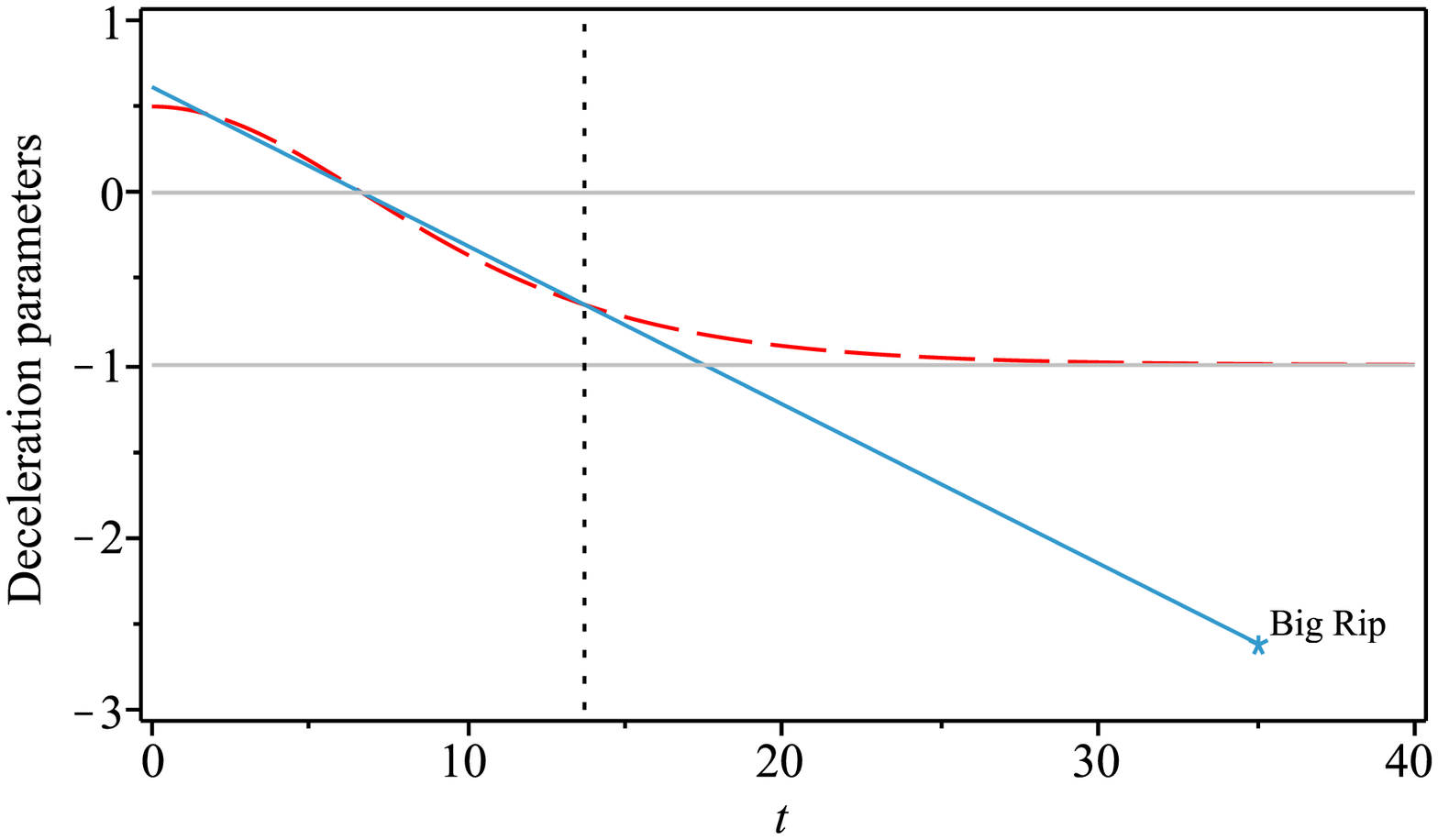}}
\end{figure}

\begin{figure}[ht]
\floatbox[{\capbeside\thisfloatsetup{capbesideposition={left,top},capbesidewidth=0.35\textwidth}}]{figure}[\FBwidth]
{\caption{The jerk parameters $j$ versus cosmic time $t$ for the LVDP (solid) and $\Lambda$CDM (dashed) models. The vertical line represents the present time of the universe $13.7$ (Gyr). The curve of the jerk parameter in the LVDP model ends at $t=34.992$ (Gyr) since the universe ends with a Big Rip at this time.}\label{fig:rho}}
{\includegraphics[width=0.6\textwidth]{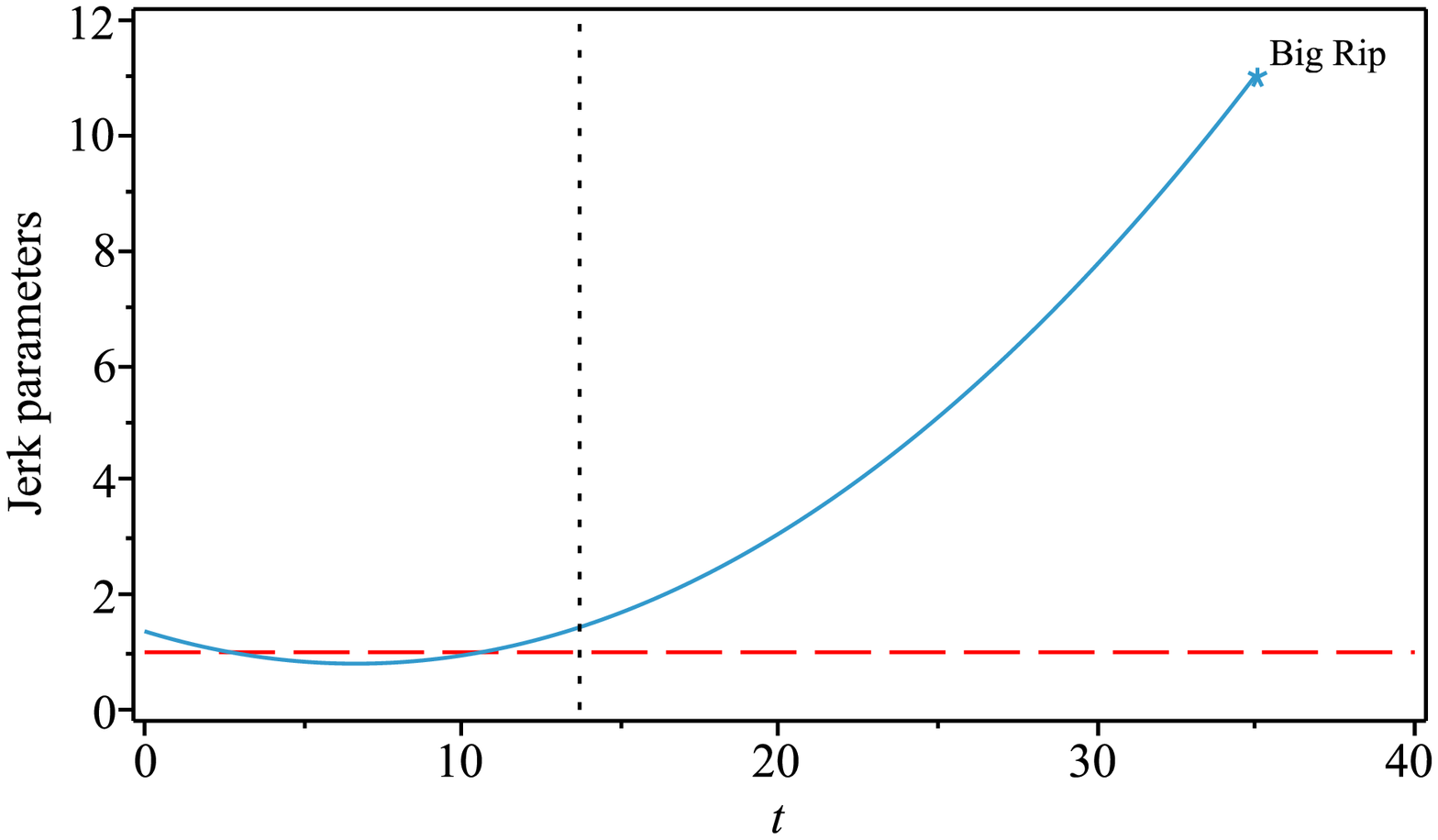}}
\end{figure}

\begin{figure}[ht]
\floatbox[{\capbeside\thisfloatsetup{capbesideposition={left,top},capbesidewidth=0.35\textwidth}}]{figure}[\FBwidth]
{\caption{The snap parameters $s$ versus cosmic time $t$ for the LVDP (solid) and $\Lambda$CDM (dashed) models. The vertical line represents the present time of the universe $13.7$ (Gyr). The curve of the snap parameter in the LVDP model ends at $t=34.992$ (Gyr) since the universe ends with a Big Rip at this time.}\label{fig:p}}
{\includegraphics[width=0.6\textwidth]{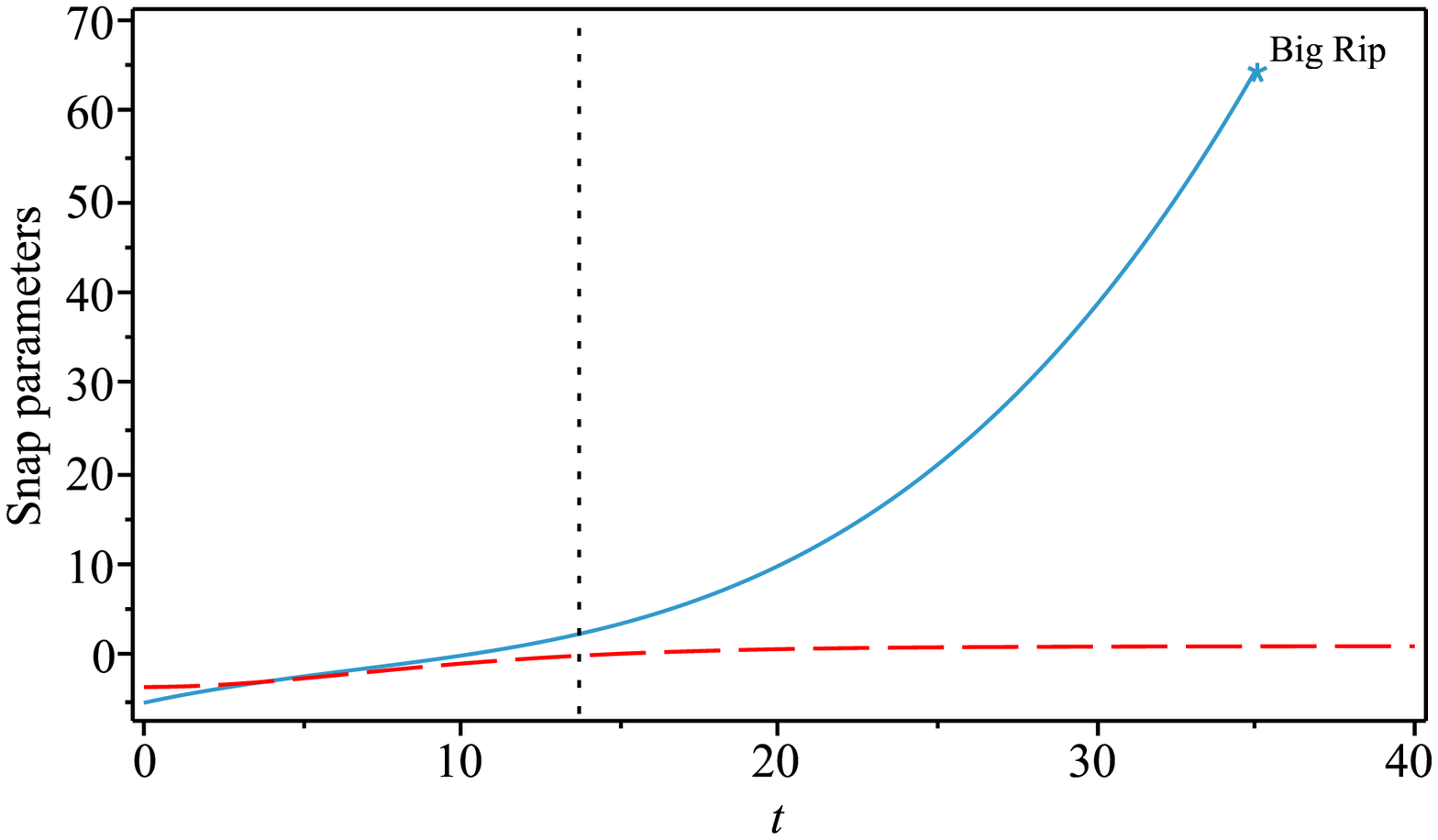}}
\end{figure}

Having obtained all the cosmological parameters both for the LVDP
and $\Lambda$CDM models, we are now able to compare them\footnote{Because our main goal in this letter is to compare the LVDP and $\Lambda$CDM models, for convenience, we regard the time parameter as dimesionless by taking $t\rightarrow \frac{t}{1\, \rm Gyr}$ and indicate Gyr in parantheses, i.e., (Gyr), to remind the reader of this.}. We choose
$q_{0}=-0.650$,  in agreement with the more recent analyses of observational results (e.g., see \cite{Komatsu11,Li11,Yun11}) of the
deceleration parameter of the present day universe $t_{0}=13.700$
(Gyr), and obtain $\Lambda=0.013$ for the $\Lambda$CDM model, which
in return implies that the universe started to accelerate at $t_{\rm
t}=6.650$ (Gyr), i.e., $t_{0}-t_{\rm t}=7.050$ (Gyr) ago.
Because these values are in good agreement with the observational
studies and we want to compare $\Lambda$CDM and LVDP models, we
simply use the above values to obtain the constants of LVDP model
such that $k=0.092$ and $m=1.613$. We can also safely set
$a_{\rm \Lambda CDM}=a_{\rm LVDP}=10$ at $t_{0}=13.700$ (Gyr) by
choosing $a_{1}=6.727$ and $a_{2}=13.144$. Determining all the
constants we now know the time evolution of all the cosmological
parameters both in $\Lambda$CDM and LVDP models. In Table
\ref{table:1} we calculate the values of all the cosmological
parameters for four different ages of the universe that are
cosmologically interesting and  help us to compare the two models.
The chosen ages are as follows: $t=t_{\rm t}=6.650$ (Gyr) when the
universe starts accelerating in both models, $t=13.700$ (Gyr) the
present-day universe in both models, $t=17.496$ (Gyr) when the universe
reaches the exponential expansion and starts super-exponential
expansion in LVDP model and $t=34.992$ (Gyr) when the universe ends with a
Big Rip in LVDP model.
\begin{table}[ht]
\caption{Values of the cosmological parameters for the $\Lambda$CDM and LVDP models at some crucial times.}
  \label{table:1}
\begin{center}
\begin{tabular}{cc|c|c|c|c|c|c|c|l}
\cline{2-9}
\multicolumn{1}{l|}{} & \multicolumn{2}{c|}{$t_{\rm t}=6.650$ (Gyr)} & \multicolumn{2}{c|}{$t_{0}=13.700$ (Gyr)} &
\multicolumn{2}{c|}{$t=17.496$ (Gyr)} & \multicolumn{2}{c|}{$t=34.992$ (Gyr)} \\
\cline{2-9}
\multicolumn{1}{l|}{} & $\Lambda$CDM & LVDP & $\Lambda$CDM & LVDP  & $\Lambda$CDM & LVDP & $\Lambda$CDM & LVDP \\
\cline{1-9}
\multicolumn{1}{|l|}{$a$} & 5.339 & 5.351 & 10.000 & 10.00 & 13.168 & 13.144 & 42.660 & $\infty$  \\
\multicolumn{1}{|l|}{$H$} & 0.114  & 0.115 & 0.075 &  0.074  & 0.070 & 0.071 & 0.066 & $\infty$  \\
\multicolumn{1}{|l|}{$q$} & 0.000 & 0.000 & -0.650 & -0.650& -0.823 & -1.000 & -0.994 & -2.613  \\
\multicolumn{1}{|l|}{$j$} & 1.000 & 0.801 & 1.000 & 1.435 & 1.000 & 2.301 & 1.000 & 11.044  \\
\multicolumn{1}{|l|}{$s$} & -2.000  & -1.602  & -0.050 & 2.346 & 0.471 & 6.204 & 0.982 & 64.488 \\
\cline{1-9}
\end{tabular}
\end{center}
\end{table}
We note that the deviations between the LVDP parameters and the
$\Lambda$CDM parameters are  negligibly small  both  during the time
of transition and today and hence during all this time interval
from the transition time to the present-day universe, since all the
parameters vary monotonically during this time interval, as would be seen from the figures.
The values of the jerk parameters are separated slightly, but remain
in agreement with  observations both for the LVDP and $\Lambda$CDM
models. The snap parameters, on the other hand, get separated more
compared to all the other cosmological parameters. However, the
definition of the snap parameter involves the fourth derivative of
the scale factor, and hence it is not an easy task to resolve
observationally any deviation between
 the snap parameter values of the above models. Therefore, we conclude that the behavior of the LVDP and $\Lambda$CDM models
 are almost indistinguishable during the observed past of the
universe. To substantiate this conclusion, we depict the scale
 factors in Fig.\ref{fig:sf}, Hubble parameters in Fig.\ref{fig:h}, deceleration parameters in Fig.\ref{fig:q}, jerk parameters in Fig.\ref{fig:rho}
 and snap parameters in Fig.\ref{fig:p} for $\Lambda$CDM (dashed lines) and for LVDP (solid line).

Any separation between the $\Lambda$CDM and LVDP models do not grow
continuously and hence the two models remain indistinguishable up to
the present age of the universe. A continuous separation between the
two models just start after the present age of the universe. Hence
both models would exhibit a similar behavior in the near future but
evolve rather differently into the far future. The LVDP model reaches the exponential expansion phase ($q_{\rm LVDP}=-1$) at $t=17.496$ (Gyr) and enters into a super-exponential expansion phase
 ($q_{\rm LVDP}<-1$), while  $q_{\rm \Lambda CDM}=-0.824$. Moreover,
 in the LVDP model the size of the space and the Hubble parameter diverge at $t=34.992$ (Gyr); while the size of the space remains finite
 and exhibits a power-law expansion with a deceleration parameter value $q_{\rm \Lambda CDM}=-0.994$ for $\Lambda$CDM at $t=34.992$ (Gyr). The divergence of the Hubble parameter in LVDP model, because the
square of the Hubble parameter $H^{2}$ is proportional to the
effective energy density of the universe in general relativity, also tells us that the
energy-density of the universe diverges for $t=34.992$ (Gyr).
 Hence, the universe ends with a Big-Rip at $t=34.992$ (Gyr) in the LVDP model, while it is still approaching the de Sitter phase in the
 $\Lambda$CDM model, where the universe is empty and only a vacuum energy (i.e., the cosmological constant)  exists.

\medskip

In conclusion, the cosmological kinematics we obtain via the LVDP
ansatz are almost indistinguishable from those of the $\Lambda$CDM
model up to the near future of our universe. Therefore, because the
$\Lambda$CDM model fits the observational data quite well all the
while, the LVDP would do so as well. The current observational data
give us information concerning the past kinematics of the universe
only and hence we do not have any reason to favor the $\Lambda$CDM
model over the LVDP model on the observational basis alone. In the LVDP model, the effective energy-momentum tensor is not introduced as a source but rather is obtained using the LVDP kinematics in the gravitational field equations. The effective source thus we obtained \cite{LVDP12} in Einstein's theory of general relativity exhibits a quintom DE-like behavior. Hence we say that the accelerated expansion in our LVDP model is driven by a quintom DE field. However, this has to be checked if one uses a generalized theory of gravity and/or includes a 'DE component' of the universe from the start. In $\Lambda$CDM model, on the other
hand, the accelerated expansion is driven by the inclusion of a
positive cosmological constant into the Friedmann equations in the
presence of a dust source. However, the presence of a cosmological
constant gives rise also to one of the most pressing conceptual
problems in physics that one may avoid in the LVDP model.

Finally, we  wish to  re-assert that the LVDP ansatz can safely be
used for generating exact cosmological models that generalize, in
particular, many of the models in the literature  already obtained
via the constant deceleration parameter ansatz, which cannot be
consistent with all observations since such models do not exhibit a
transition from a decelerating expansion to an accelerating
expansion, whereas LVDP ansatz does this consistently with
observations. Furthermore, the LVDP ansatz is a good candidate for
studying cosmological models with  Big Rip futures and yet remain
consistent with the present-day observations.

\bigskip

\noindent  {\large \textbf{Acknowledgement}}

\medskip

\noindent The authors appreciate the financial support received from
the Turkish Academy of Sciences (T\"{U}BA).

\newpage


\end{document}